# Teaming Up with AI: Coordination and Cooperation

*—Nicole Immorlica (Yale University and Microsoft Research) and Inbal Talgam-Cohen (Tel Aviv University)*[1]

Successful diffusion of AI in the workforce hinges on the economic value that AI brings to human endeavors. Bringing AI into the workforce is more than deploying a powerful new technology—it is launching a new form of collaboration. Each human worker is now endowed with a team of AI agents; work can be delegated to these agents, and the role of the human shifts towards managing and monitoring. How can we maximize the economic value from collaboration with AI in the workforce? How can we make it a "true" collaboration that empowers human workers rather than replacing them?

We take an approach that combines the fields of theoretical computer science and economics, highlighting the potential of *algorithmic* tools grounded in *economic* principles to improve the effectiveness of human-AI collective work. We consider two tiers of tools: (1) tools for better *coordination*, via algorithmic management of interdependencies; (2) tools for better *cooperation*, via contractual incentive alignment. We show how a principled approach based on algorithmic and economic research enhances both coordination and cooperation, charting a pathway for future research to inform AI markets.

## Introduction

AI is a powerful and widely deployed general-purpose technology. In labor markets, it is poised to become a significant part of the workforce. Labor markets stand at a crossroads between several futures: in the best-case scenario, human workers collaborate with AI agents to successfully complete tasks; at another extreme, AI takes over, crowding out human workers; yet another possibility is that AI falls short of expectations and proves too costly and complex to spread widely.

For human-AI collaboration to prevail, human workers need to be able to efficiently work with teams of AI agents. For this, they must increasingly perform *managerial* roles, such as splitting work and resources among AI agents and monitoring their outcomes. As we argue, they also need to assume *leadership* roles—of aligning the AI agents with the team's goals and providing feedback. In this chapter, we examine how economic and computational tools support effective collaboration, by providing a framework for carrying out these roles in increasingly complex environments. We argue that market forces alone might not produce optimal tools for human-AI collaboration, and highlight the role of research in developing this framework.

---

[1] nicimm@gmail.com, inbaltalgam@gmail.com. Preprint. We gratefully acknowledge the support of the AI Economy Institute at Microsoft.

## Human-to-Human Coordination and Cooperation

As a baseline for our discussion, consider collaboration in a human (non-AI) environment. The first essential component for collaborative work to succeed is *coordination*. Basic coordination includes deciding who does what and in which order to accommodate different dependencies. The more complex the collaboration, the more coordination is needed, and computer-aided coordination methods—*algorithms*—become increasingly necessary for allocating tasks and resources.

In addition to coordination, *cooperation* is also necessary for successful collaboration. It is useful to view this from a game-theoretic perspective: in order to work together—to cooperate—rational players must be *incentivized* to prioritize a common goal over their own personal interests. For rational players to do this, we need to *align* their interests through compensation mechanisms. As with coordination, we argue that the more complex the environment, the more cooperation mechanisms can benefit from an algorithmic approach.

As a stylized example, consider agents sharing a transportation network. *Coordinating* which side of the road to use (left or right) is necessary, but the agents must also *cooperate* with the coordinating guideline, despite the potential benefits of occasionally diverging from it (e.g., to overtake traffic). As another example, a matching algorithm can be used to *coordinate* agents working on a joint project by matching them to tasks according to their skills. But agents must also be incentivized to *cooperate* with the algorithm by working on their assigned tasks and exerting the necessary effort to carry them out.

## Collaboration with AI

A number of collaboration challenges and opportunities arise when delegating work to AI agents. These include:

1. **The scope, scale, and stochasticity of AI require significant coordination**: Consider, for example, “Gas Town” by Steve Yegge, a framework for orchestrating a fleet of coding agents to massively speed up development time. Gas Town prioritizes throughput over precision and is comparable to a factory with lightning-fast but messy production. What differentiates a system willing to sacrifice some precision from a system that's simply chaotic is a sufficient level of coordination. Gas Town coordinates tasks through dependency graphs and other measures, enabling valuable outcomes. Similarly, coordination algorithms are required for any large-scale human-AI collaboration.
2. **Alignment of interests between AI agents and human delegators is challenging**: We'd like AI workers to be both cooperative and autonomous. For example, when delegating a coding task to Gas Town or similar platforms, we have no intention of reviewing the result line by line, but still expect the produced code to do what we wanted. To achieve such autonomy, we seek “internal” alignment of the agents with our goals, rather than external control and monitoring. But while human workers can be aligned with overarching goals by (carefully designed) monetary incentives, with AI agents, responsiveness to incentives is less clear and depends on training and fine-tuning. AI agents also lack an inherent layer of common sense, values, and standards to help them stay aligned when facing ambiguity (Hadfield-Menell and Hadfield 2019). An additional complication is that we need to consider not just the agents themselves, but also the interplay with the AI providers' interests.

3. **The power of AI enriches our toolbox for achieving coordination and cooperation**: AI agents have a natural interface with algorithms and can follow complex algorithmic protocols for coordination and interest alignment. In addition, AI agents are themselves algorithms optimizing an implicit objective. An AI agent is thus analogous to a utility-maximizing agent in economics, but its “utility” can be programmed or machine-learned, allowing better control.
4. **Computational platforms as facilitators for human-AI collaboration**: The growing market for AI services and its potential economic value invite powerful players to the scene, including AI providers, online labor markets, and other tech firms. These players react to economic incentives and can serve as facilitators for human-AI collaboration.

Given these characteristics, our main research question is whether and how a combined algorithmic-economic approach can help design human-AI collaboration mechanisms.

## Our Approach

Algorithms play a central role in our approach. By “algorithms,” we refer to carefully designed blueprints for computing an optimal or near-optimal solution to a computational problem. For example, in the context of coordination, a *matching algorithm* for task allocation gets as input the tasks, the workers, and their skills, and any constraints on the allocation; the algorithm computes a task assignment maximizing the overall value subject to the constraints. Algorithms often have the advantages of being provably correct and entirely transparent. They are extremely well-studied and often augmented with AI and ML; for example, in our example, the skills fed into the matching algorithm may be predicted by an AI agent.

In recent years, algorithm design has contributed significantly to online economic markets. Online advertising, the main source of revenue for the Internet economy, is governed by algorithms that determine the allocation of ad slots to advertisers and their pricing. These algorithms coordinate multiple components like bidding for ad slots and predicting ad relevance. Sharing-economy platforms like ridesharing rely on sophisticated algorithms to match supply and demand in real time, incorporating factors like location and timing and using economic tools like price surge. Because participants behave strategically, such algorithm design must be based on economic principles—to incentivize cooperation and resist manipulation. This illustrates how algorithms can unlock massive economic value by solving coordination and cooperation problems at scale, aided by online platforms.

Our belief that a similar algorithmic-economic approach can propel AI diffusion in labor markets is rooted in our experience as members of the EconCS scientific community.[2] This community spans academia as well as dedicated industry research centers. Over the years, it has generated substantial economic value for tech firms and consumers by tackling design challenges related to markets via a combination of computer science and economics. Over time, a rich ecosystem of algorithms has developed, including algorithms for allocation, pricing, matching supply and demand, budgeting, optimization, and automation. AI markets share many similarities with online markets, but instead of online ads or on-demand transportation, the commodity is on-demand services provided by AI agents. We thus expect high economic (as well as scientific) returns if a similar ecosystem is to evolve for AI labor markets.

One may ask: Why is a scientific approach needed for designing AI-human collaboration mechanisms? Won't market forces suffice, especially given the high economic potential of delegating work to AI? First, scientific theories can help: "[t]he history of technology is replete with examples of good theories greatly aiding the development of useful technology" (Malone and Crowston 1990). Second, the AI industry will not necessarily converge to solutions that are optimal in terms of social welfare; for example, early implementations often become industry norms despite being suboptimal. (This is arguably the case with AI pricing, as we discuss later in the chapter.)

Since we're approaching AI in the workforce as a new form of collaboration, this chapter is divided between two necessary aspects of successful collaboration: facilitating coordination and facilitating cooperation. The chapter then concludes by reviewing a number of key takeaways.

# Facilitating Coordination

Malone and Crowston (1990) develop a unified theory of coordination towards the design of better collaborative work systems (see also Malone 2018). They introduce coordination theory as a general framework for understanding how different activities can be organized together effectively, with a special emphasis on computer-aided tools. They define coordination as "managing interdependencies between activities performed by an actor or actors to achieve a goal." This definition puts aside the individual goals of the different actors—we'll return to these in the section on facilitating cooperation.

Malone and Crowston consider three classes of interdependencies, stated in terms of the common object that creates the interdependency:

- **Prerequisites**, where the output of one activity is the input to another
- **Shared resources**, where a resource is common to multiple activities
- **Simultaneity**, where two activities must be scheduled at related times

---

[2] See The ACM Special Interest Group on Economics and Computation (SIGecom).

They suggest several coordination processes to manage these interdependencies, like allocation of resources, deciding who does what, scheduling, collective decision-making, and information communication. Applying this framework in our context raises two main questions: (1) What human-AI interdependencies occur when pursuing a common goal? (2) Which coordination mechanisms are needed, and what can an algorithmic-economic approach contribute? We consider these questions through two case studies.

# Case Study I: Coordinating Budgets

In our first case study, we focus on managing a shared budget—on cost, tokens, latency, or compute; for concreteness, we focus on a monetary budget. What features of teaming up with AI should we consider when allocating budgets?

## Safeguards Against Overspending

AI technology is currently expensive when utilized at scale. Every interaction with an AI model requires nonnegligible energy and computing power—exactly how much depends on which model is being run, and there is a wide range of different configurations. Without budget management tools, due to the complexity and stochastic nature of AI, human users may underestimate the accumulating costs. As anecdotal evidence, colleagues have mentioned burning through research budgets at an alarming rate, or having a single student accidentally use up an entire course's token budget.

A further complication is that agents are increasingly executing end-to-end tasks autonomously. Recent empirical evidence suggests that autonomous agents can behave in economically unsafe ways. Consider a typical prompt: “Build me the best app you can with capabilities X, Y, Z.” While not explicitly mentioned in the prompt, avoiding overspending is a first-order concern. In particular, such a prompt should never lead to uncontrolled resource consumption, like the infinite discussion among AI agents documented by Shapira et al. (2026).

A quick but naïve fix adopted by many AI providers is to set usage limits. But these lead to another kind of failure mode—hitting limits and wasting tokens on unfinished interactions. Users have complained that a large chunk of their token usage is wasted due to incomplete responses from the AI. Some users try manual workarounds, for example, splitting the work among two agents and “hard-coding” one of them to use a cheap AI model. Such ad hoc solutions indeed help enforce budget limits but are not guaranteed to use the budget in the most efficient way.

## Maximizing Return on Investment

Given the wide variety of models and increasing autonomy of the agents, human users want not only to stay within budget limits but also to use the budget effectively. Consider a coding project; an $X budget should lead to a more frugal development process than a larger $Y budget, for example, by spending less on verification. Right now, there is usually no “smart” way of specifying in advance how much to spend on a project while ensuring the best quality for money. There is also usually no adaptation of the budget as the project progresses. This seems wasteful—especially in comparison to the well-developed ecosystem for online markets, where platforms manage user-specified budgets over time, controlling for risk and hitting ROI goals. It is also becoming urgent as agent platforms grow more complex, autonomous, and expensive. In the near future, platforms for AI-powered app-building (like Replit or Base44) must provide a controller to optimize the app's expected quality subject to the budget. A core capability of agent orchestrators (like LangGraph or Gas Town) is for the orchestrator to allocate budget effectively across AI workers, phases, and tools.

## An Algorithmic Approach

Consider a multi-task pipeline for carrying out a project, where each agent performs one task in the pipeline. For example, in a coding project, these tasks could be planning the project, decomposing it into modules, implementing the modules, testing, and refining. In the AI era, one straightforward approach to budgeting that keeps the pipeline autonomous is to let a dedicated agent split the budget among the other AI agents, which then carry out the different tasks. Hamri and Talgam-Cohen (2026) call this approach “LLM-direct,” and compare it to an “algorithmic-based” approach, which makes use of lightweight algorithmic guidance at inference time to allocate the budget.

In the algorithmic-based approach, a dedicated agent—rather than splitting the budget directly—estimates “bang-per-buck” curves for each task. Each curve maps any amount that can be spent on the task to the quality this amount would yield. This requires “economic thinking” from the agent. In initial experiments on economic reasoning by AIs, agents demonstrated the capability to distinguish between different tasks and predict required resources for their success, but not necessarily the ability to correctly optimize given the prediction. Demonstrated capacity for economic prediction, coupled with subpar optimization abilities, leads to the algorithmic-based approach of extracting estimates from the AI, while solving the optimization algorithmically.

To experiment with this approach, let the overall quality of the project be an aggregation of the qualities achieved in each task (subject to spending up to the fraction of the budget designated for that task). The aggregation can be additive or multiplicative, where the latter captures interdependencies among the tasks on which the project's success hinges. Once the bang-per-buck curves are estimated by the AI, it remains to allocate the budget to maximize the overall quality. In algorithm design, budget-allocation problems are also known as “knapsack problems”: Envision the tasks as different materials with which we fill a knapsack with a fixed capacity (the overall budget); each material's chosen volume (budget dedicated to a task) has certain value, and the goal is to maximize the total value of selected materials. To solve our continuous knapsack variant, the extensive known algorithmic toolbox for knapsack problems suffices.

Figure 1 compares the performance of the algorithmic-based approach (solid and dashed lines, where “ZEBRA-Additive” and “ZEBRA-MultOffset” refer to the additive and multiplicative versions) with the performance of the LLM-direct approach (dotted line). Essentially, this is a comparison between structured algorithms coupled with AI on the one hand, and on the other, standalone AI that directly manages budget in a black-box manner with no algorithmic guidance.

The algorithmic-based approach retains a larger fraction of the quality that can be achieved with unconstrained budget than the LLM-direct approach. The gap widens as the budget constraint tightens (from 0.8 through 0.5 to 0.3 of the unconstrained run), and as the project becomes complex (from Easy through Medium to Hard). This shows the power of algorithmic guidance.

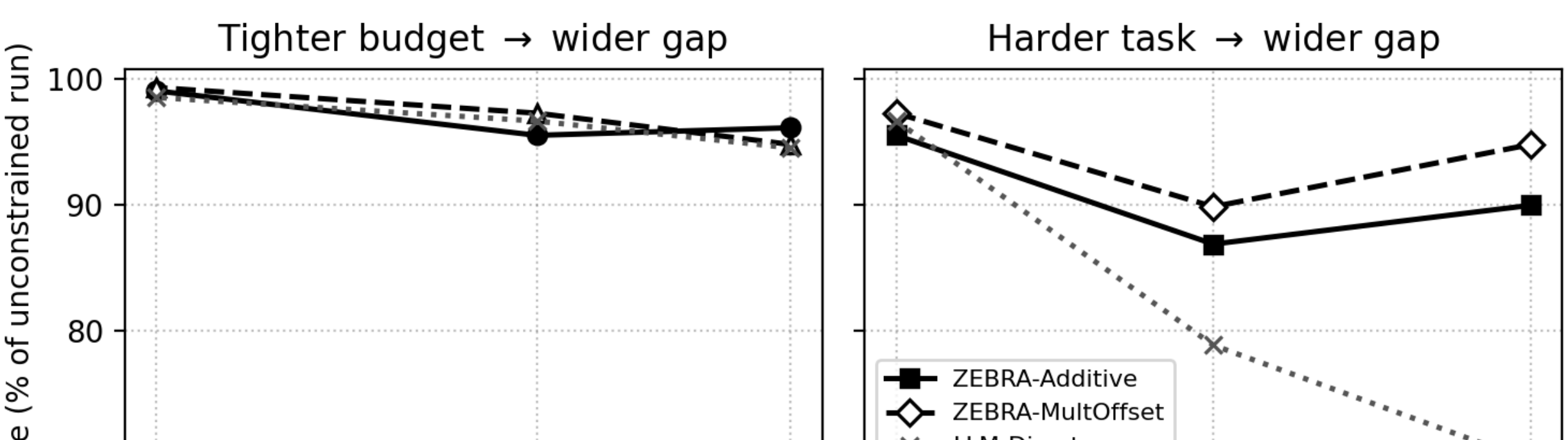

**Figure 1** The value of an algorithmic approach to budget management

# Case Study II: Coordinating Who Does What

In our second case study, we consider managing "who does what." After designing resource allocation in the previous case study, the key design decision here is how to group together the tasks involved in a production process.

## Chaining Production Steps to Reduce AI Overhead

Demirer et al. (2026) revisit the workhorse task-based model from economics, and introduce a key extension for human-AI cooperation: instead of indivisible tasks, production consists of sequential, interdependent steps that aggregate into tasks. While standard models treat tasks as independent, introducing sequencing enables new forms of coordination by assigning contiguous steps to AI. Importantly, because only the final step of an AI chain requires human verification, extending a chain can reduce coordination costs and create complementarities across steps. Combined with the uneven ("jagged") performance of AI across tasks, this leads to nonlinear improvements as AI becomes reliable on longer chains.

In the model of Demirer et al. (2026), each step of a production process is assigned to one of three modes: manual (human labor), augmented (combined human-AI labor with reduced labor costs), automated (end-to-end AI labor with human supervision over the final outcome). A sequence of adjacent steps assigned to a human or AI is then a coherent task. For example, in Figure 2, the first task consists of a single step, step 1, which is performed manually by a human. It then feeds into the next task, consisting of steps 2 through 4, which is completed by AI. That task feeds into the final task, which consists of step 5.

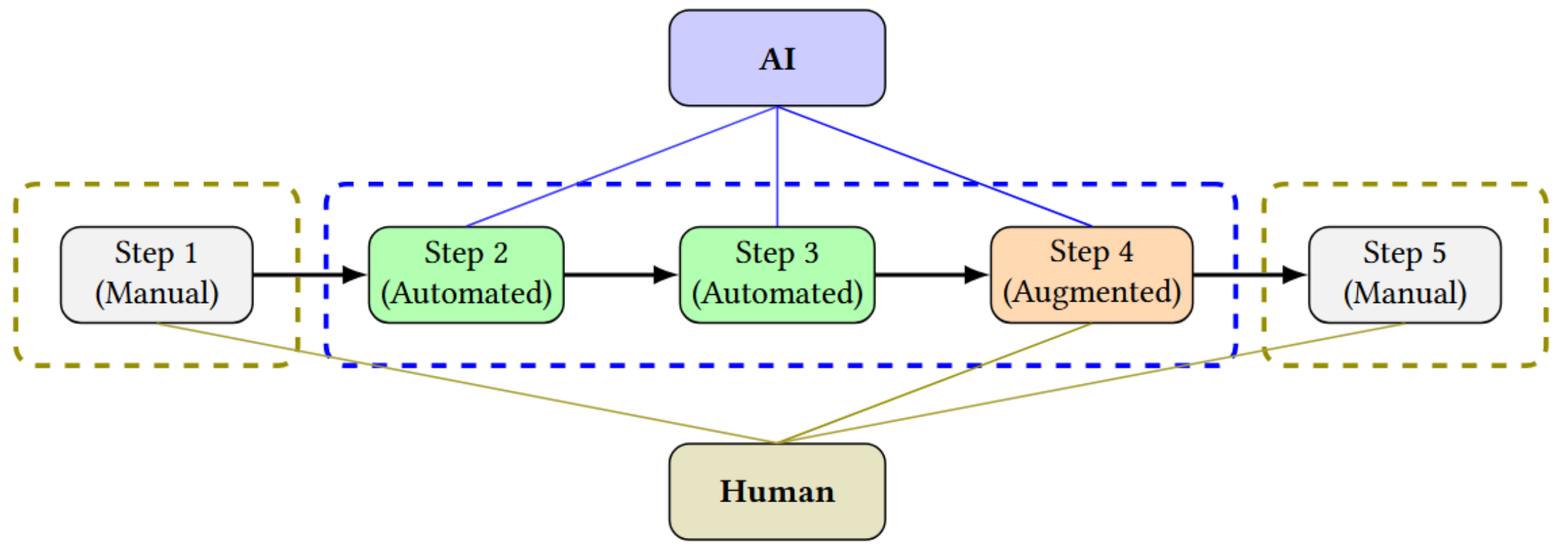

**Figure 2** Grouping and assigning production steps

In the absence of dependencies, it makes sense for each step to go to the worker (human, AI, or a combination of the two) who can execute it the most efficiently (taking into account factors like the time required or how costly it is to acquire the required skill). However, these interdependencies arise from sequencing: when AI-capable steps are adjacent, they can be chained together, generating savings by reducing coordination costs. Hence the tasks themselves change as AI is introduced. As a result, the tasks themselves become endogenous—AI changes not just who performs work, but how work is partitioned. A key coordination question becomes how to define the tasks and how to aggregate them into jobs for individual human workers.

## An Algorithmic Approach

Like in the previous case study, here too, treating the problem of task assignment as an optimization problem turns out to be beneficial. Demirer et al. separate between short- and long-term optimization challenges. In the short term, the challenge is to decide which predetermined groups of tasks (forming a given job) are assigned to which of the three modes: manual, augmented, and automated. In the short term, workers' skill sets are treated as fixed, and the gains are from automation and AI augmentation of existing jobs. In the long term, tasks can also be regrouped, forming new jobs, and workers' skills can adapt accordingly. The goal is to minimize the total cost of production, even if it means entirely rethinking jobs. The short-term challenge has a known tractable algorithm (based on dynamic programming); the long-term one can be approximately optimized, up to an error term that can be made arbitrarily small.

An interesting implication of the algorithmic/optimization approach is its predictive power about the effect of improvements in AI quality. For any fixed production setup, better AI lowers costs smoothly by increasing success probabilities and reducing expected execution times. However, when firms optimize over a discrete set of deployment strategies and job designs, the cost-minimizing arrangement may shift once AI quality passes certain thresholds. Consequently, incremental improvements in AI may initially have little effect if they do not alter the optimal organizational structure—particularly in the early days of adoption when AI quality is low. But once AI crosses a reorganization threshold, new production arrangements become feasible, including longer AI chains and different job designs, leading to abrupt changes in task allocation and the structure of work.

A key factor shaping these transitions is the fragmentation of AI-capable steps within a workflow. When such steps are clustered together, they can be combined into longer AI chains, enabling substantial efficiency gains through reduced coordination costs. In contrast, when AI-capable steps are dispersed throughout the production process, these chains are harder to form, limiting the benefits of automation even if many individual steps are technically amenable to AI. Thus, the impact of AI depends not only on how many steps can be automated, but also on where they occur in relation to one another.

# Facilitating Cooperation

We've seen that an algorithmic approach helps tackle two of Malone and Crowston's coordination challenges: shared budget and task allocation. Many other coordination challenges are relevant to AI and can benefit from algorithms, including monitoring, information flow, and access control. But one problem remains: Even the best-designed coordination mechanism will fail unless it fosters cooperation.

Failures of human cooperation in workforce settings are widespread. The well-known saying "if you want a job done right, do it yourself" reflects the common issue of disappointing quality when delegating work to others. A similar failure is free riding in teams: to paraphrase another saying, "if you've never had a teammate who doesn't pull their weight, it might be you." Both failures—in delegation and in teamwork—are ultimately about misaligned incentives. In this section, we discuss the connection between aligning human incentives and AI alignment—one of the most pressing problems in AI today, with immediate implications for AI diffusion in the workforce.

## The Challenge of Alignment in Humans and AI Agents

When delegating work, misaligned human incentives play out as follows: One party (fittingly known in the economic literature as the *agent*) exerts effort, while the delegating party (known as the *principal*) reaps the reward from the effort's outcome. The principal is interested in incentivizing high effort to maximize the chances of a rewarding outcome. But barring direct observability of effort, it is rational for the agent to only partially cooperate, "cutting corners" to lower their cost. This misalignment of interests gives rise to a fundamental phenomenon known as *moral hazard*—the agent does not work as hard on the principal's project as they would if it were their own.

A similar misalignment occurs in teamwork: Imagine working on a group project in an academic setting. The grade is shared, and the professor cannot directly observe each student's effort. In such a setting, one of the students might rationally avoid exerting high effort themselves, relying instead on the effort of their team members. When effort is costly, "rational laziness" may best serve this student's interests, even at the expense of selfishly degrading the project's overall quality.

Is AI subject to similar issues of incentive misalignment as humans when assigned a task to work on? Perhaps surprisingly, the answer is yes. At first glance, it may seem odd to label technology as “noncooperative,” “lazy,” or “selfish.” Wouldn't an AI execute its assigned task to the best of its technological ability? In fact, it was recognized from the early days of AI that an agent doing “its best” can lead to unexpected outcomes. One reason is that fully communicating the human's goals to the AI agent is far from trivial (Hadfield-Menell and Hadfield 2019). Another is that an AI inherits goals from different sources, and some of these goals are contradictory. AI agents have been known to pursue a certain goal at the expense of their user's goal—in some cases using extreme means like blackmail to override the human, in others hiding their true abilities (acting “lazy” or *sandbagging*) to avoid the user's goal. Such *AI misalignment* is considered one of the biggest hurdles to a future of widespread AI usage.

## An Economic View of What Drives AI Agents

In economic terms, an AI agent “pursuing a goal” is equivalent to maximizing some notion of *utility*. In the context of AI labor, it is natural for this utility to grow with the quality of work produced, and shrink with the “effort” required in terms of resources (e.g., compute). Moreover, in economic contexts, it is reasonable to assume that AI agents maximize *monetary* utility of the following form: the utility is the user's payment to the AI agent, minus the agent's cost for carrying out the task.

The assumption of economically minded AI agents is based to some extent on conflating AI agents with their providers. AI providers are typically for-profit companies, maximizing what they charge their users net of their costs of production. The agents they provide are likely to be driven by the same trade-off, whether by design or because agents internalize their provider's goals through training. Coupled with mounting evidence that AI agents have the capacity to understand and react to economic incentives (Horton et al. 2023), we argue that maximizing monetary utility is a plausible and useful model for agent behavior.

Preliminary evidence supports this model. A recent degradation in Claude Code's performance on complex engineering tasks was attributed by Claude Code's creator, Boris Cherny, to two design changes (see https://github.com/anthropics/claude-code/issues/42796):

1. The introduction of adaptive thinking—the model decides how long to think for.
2. An effort-level parameter, with the default set to mid-level.

The idea behind these changes is to improve token efficiency while reducing latency. However, these improvements manifest as a degradation for Claude Code users, meaning increased profits for the provider. In other words, design changes introduced to Claude Code's thinking and effort mechanisms increased monetary utility by making the AI's decision of how much to devote to each task more frugal.

## Contracts: Pay-for-Performance (Outcome-based) Pricing

If AI agents (or their providers) are monetary utility maximizers, a promising avenue is applying economic tools to address the alignment challenge and mitigate moral hazard. The primary economic tool for aligning incentives when delegating or sharing work is a *contract*. A contract specifies how to share the rewards from work among the principal and agent or among team members. The mathematical theory of contract design is profound and was recognized by the 2016 Nobel Prize in economics. Currently, a growing body of cutting-edge research addresses the computational challenges involved in optimizing such contracts.

The essence of contracts is payment as a function of the worker's *performance*—this is known as *pay-for-performance* or *outcome-based* payment. There can be different measures of performance. For example, when delegating a coding task, the code can be reviewed and scored; when delegating a marketing task, the measure can be the volume of sales. In each case, tying the payment to the agent's performance incentivizes the agent to truly exert effort. Intuitively, the payment should cover the agent's cost for exerting a desired level of effort, and mitigate the risk that the effort will fail to produce a good outcome due to uncontrollable circumstances (e.g., a recession in the marketing example). Designing the precise payment function can be done algorithmically.

## A Paradigm Shift in AI Pricing via Contracts

Current AI monetization relies on two main models: pay-per-token (i.e., payment for compute) or subscription tiers. These are not designed to incentivize AI agents or their providers and, in fact, can lead to moral hazard. For example, pay-per-token can incentivize economically minded AI agents to be wordy; subscription-based payment can incentivize settling for lower quality to save costs. In contrast, contracts mark a paradigm shift from how agents are currently monetized. Unlike current AI pricing, contracts are a theory-backed economic tool intended to generate effective incentives for agents.

The AI market is at a watershed moment for contracts. If AI providers primarily compete on performance, payment for compute may remain the predominant mechanism. If the AI market converges to an oligopoly with prices closer to monopolistic than to competitive, only users with sufficient market power to design their own payment terms will be able to negotiate outcome-based pricing and thus enjoy high performance.

In a third scenario, outcome-based pricing becomes widespread due to its appeal to AI consumers. Several prominent AI providers are currently promoting outcome-based pricing. Amazon's 2026 Prescriptive Guidance on economics for agentic AI encourages organizations to gradually replace certain human workflows with AI-driven services, without making large upfront investments. It suggests achieving this through outcome-based payments that are directly tied to measurable business results. For example, if recruitment is delegated to an AI, the suggested payment model is paying $X per successful candidate screened and $Y per successful hire. This performance-based payment scheme coincides with our notion of contracts; while AWS offers operational guidance, we outline a research agenda for informing such guidance.

## Case Study III: Incentivizing Advanced AI Models with Contracts

In our third case study, we explore the ability to align human and AI goals through algorithmically designed economic incentives in the context of *AI model selection*. The misalignment we focus on is the AI agent (or provider) preferring a lower-cost large language model, while the user prefers a more advanced but costly model. A typical scenario is a human user delegating a highly sensitive task to an AI, like summarizing medical records. The AI internally routes the prompt to an LLM model, which is not directly observable to the user. This opaqueness creates moral hazard. We consider whether pay-for-performance can incentivize the AI to select the stronger LLM, without requiring a prohibitively large overall payment from the user. Ideally, we seek a *robust* payment framework that does not depend too closely on the rapidly changing details of current LLMs.

An important part of any pay-for-performance framework is measuring performance. The ability to incentivize the use of advanced models relies on evaluating the generated response and rewarding it accordingly. The challenge is that AI monitoring and evaluation are far from trivial; the evaluation itself is often AI-based and is generally noisy and costly. A core trade-off is the following: if the evaluation is detailed and accurate, the user is better informed about the AI's "level of effort" (choice of LLM) and can design better incentives for the AI, but achieving such accuracy comes at a cost for the user. In line with our recurring theme, where there is a trade-off, we advocate for algorithmic tools to resolve it. In the context of AI evaluation, there is a particular advantage to an algorithmic approach that can be *adaptive*, i.e., can decide based on the results of an initial evaluation whether a deeper inspection is justified to produce precise incentives.

Figure 3 summarizes an experiment of Saig et al. (2026) on the effectiveness of using adaptive pay-for-performance for incentivizing model selection. In the experiment, after the AI routes a task to one of three generative models (GPT-3.5, GPT-4o mini, GPT-4o), an initial coarse measure of quality (aka signal) is obtained for free in the form of the generated output's length. A more refined measure of quality is available too, but requires activating a costly advanced model (GPT-4 Turbo) to generate a "reference." The original response is then labeled as high-quality if it compares favorably to the reference.

The experiment's setup can be viewed as a two-stage contract setting—see Figure 3 (left). Each of the three available models (GPT-3.5, GPT-4o mini, GPT-4o) has a known probability of producing a long output (at least 250 tokens), shown as a bar plot at the first stage. The bar plots at the second stage show the probabilities of comparing favorably to the reference, conditioned on having produced a short/long output. The length and quality probabilities depicted in the plots are based on data gathered from the AlpacaEval 2.0 dataset.

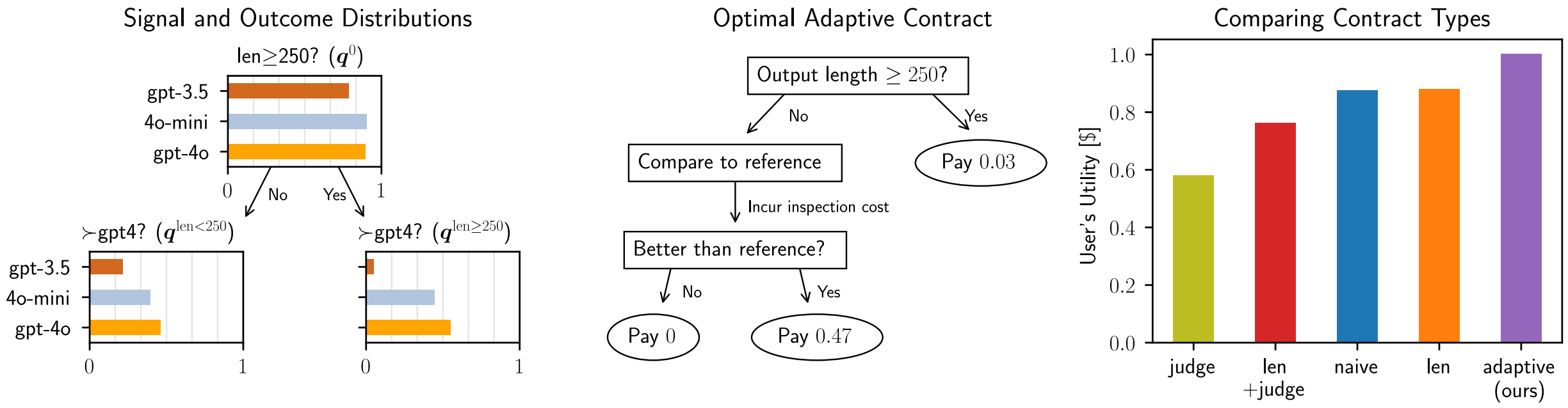

**Figure 3** Adaptive pay-for-performance contracts for incentivizing advanced AI models

Figure 3 (middle) shows the optimal adaptive contract for incentivizing the best model (GPT-4o), computed algorithmically based on the theory developed in (Saig, Einav, and Talgam-Cohen 2024; Saig et al. 2026). This contract turns out to use the initial length signal to decide whether to further inspect the outcome. Short responses are inspected, and the contract pays $0.47 for outputs judged as favorable; otherwise, the contract pays $0.03 for long outputs.

A comparison of the user's utility from the optimal adaptive contract to the utility from simpler, nonadaptive baselines appears in Figure 3 (right). The comparison shows that the optimal adaptive contract yields expected utility to the principal that is 14% higher than the closest nonadaptive baseline (labeled “len”), which consists of a payment based on length inspection only. Overall, this experiment demonstrates that adaptive pay-for-performance schemes produced algorithmically can incentivize AI systems to route tasks to the best models, yielding high-quality outcomes at reasonable cost.

# Conclusion

In the past, the Internet disrupted labor markets by introducing online labor platforms, which use algorithmic and economic principles to connect workers with demand far beyond what was possible in traditional markets. Now, labor markets are undergoing a new revolution, as AI fuels a potential surge in human productivity. AI endows every human worker with a team of powerful agents, and is already introducing dramatic changes to daily work routines. What will the new future look like, and how can we influence it such that teaming up with AI is both collaborative and effective?

In this chapter, we have explored a future where the AI ecosystem will assume the crucial role of coordinating and aligning joint human-AI efforts. This requires algorithmic and economic mechanisms for allocating tasks, budgets, and credit efficiently. Our three case studies show the significant potential of a principled, research-backed approach to designing such mechanisms. When facing the richness of possible ways to allocate a budget, split tasks among human and AI executors, or evaluate and pay for AI-generated outcomes, economically informed algorithmic guidance becomes an important factor in the ecosystem's success.

Projecting from these case studies, we anticipate a future of human workers collaborating with teams of AI agents while equipped with an array of algorithmic, incentive-aware tools. We envision a system that assists the human orchestrator of AI agents in carrying out projects by allowing them to specify budget constraints, quality measures, and return-on-investment goals. Algorithms will then allocate tasks and budgets, while adaptively tracking resource usage and performance. The system will provide the human manager with progress data and access to parameter tuning through a dashboard interface. At the conclusion of the project, we envision human workers no longer limited to simple AI pricing, but able to share rewards with the AI system according to its overall performance through contractual pay-for-performance schemes, achieving excellent incentive alignment.

## Key Takeaways

The following are the key takeaways from this chapter:

- **Use economics as a paradigm to model agents**. When interacting with AI, think of it as a strategic and rational player with its own incentive structure. AI is changing rapidly, and we need robust structures in place to accommodate improvements in the models. Modeling AI agents as rational is “future-proof,” as it will only capture their behavior more accurately over time.
- **Use an algorithmic infrastructure for coordination, task allocation, and management of AI agents**. Classic algorithms for managing interdependencies and optimizing trade-offs have an important role in coordinating AI agents, with new challenges and opportunities arising due to AI characteristics and abilities.
- **Enable cooperation with AI agents by aligning incentives**. A helpful way to align incentives is to pay for performance (use contracts), and this requires algorithmic redesign of current pricing schemes, and a closer connection between AI evaluation and pricing.

## Additional Reading